\documentclass[]{spie}  

\usepackage[]{graphicx}
\usepackage{color}

\title{SPIRou: the near-infrared spectropolarimeter/high-precision velocimeter for the Canada-France-Hawaii telescope} 

\author{\'Etienne Artigau\supit{a}, Driss Kouach\supit{b}, Jean-Fran\c cois Donati\supit{b},  Ren\'e Doyon\supit{a}, Xavier Delfosse\supit{c},  S\'ebastien Baratchart\supit{b}, Marielle Lacombe\supit{b}, Claire Moutou\supit{d}, Patrick Rabou\supit{c},  Laurent P. Par\`es\supit{b}, Yoan Micheau\supit{b},  Simon Thibault\supit{e}, Vladimir A. Reshetov\supit{f}, Bruno Dubois \supit{b},  Olivier Hernandez\supit{a}, Philippe Vall\'ee\supit{a}, Shiang-Yu Wang\supit{g}, Fran\c cois Dolon\supit{b}, Francesco A. Pepe\supit{h}, Fran\c cois Bouchy\supit{i}, Nicolas Striebig\supit{b}, Fran\c cois H\'enault\supit{c}, David Loop\supit{f}, Leslie Saddlemyer\supit{f}, Gregory Barrick\supit{d}, Tom Vermeulen\supit{d},  Michel Dupieux\supit{b}, Guillaume H\'ebrard\supit{j}, Isabelle Boisse\supit{k}, Eder Martioli\supit{l}, Silvia H. P. Alencar\supit{m}, Jos\'e-Diaz do Nascimento\supit{n} \& Pedro Figueira\supit{k}
\skiplinehalf
\begin{small}
\supit{a}D\'epartement de physique and Observatoire du Mont-M\'egantic, Universit\'e de Montr\'eal, Montr\'eal H3C 3J7, Canada; \\
\supit{b}IRAP-UMR 5277, CNR and Universit\'e de Toulouse, 14 Av. E. Belin, F-31400 Toulouse, France;\\
\supit{c}Institut de Plan\'eologie et d'Astrophysique de Grenoble, UMR 5274 CNRS, Universit\'e Joseph Fourier, BP 53, 38041 Grenoble Cedex 9, France;\\
\supit{d}CFHT Corporation, 65-1238 Mamalahoa Hwy Kamuela, Hawaii 96743, USA;\\
\supit{e}D\'epartement de physique, g\'enie physique et optique and COPL, Universit\'e Laval, Qu\'ebec, Qu\'ebec, Canada;\\
\supit{f}National Research Council of Canada Herzberg, 5071 West Saanich Road, Victoria V9E 2E7, Canada;\\
\supit{g}Institute of Astronomy and Astrophysics, National Taiwan Univ., Taiwan;\\
\supit{h}Observatoire Astronomique de l'Universit\'e de Gen\`eve, 51 Ch. des Maillettes, 1290 Sauverny, Versoix, Switzerland;\\
\supit{i}Aix Marseille Universit\'e, CNRS, LAM (Laboratoire d'Astrophysique de Marseille) UMR 7326, 13388 Marseille, France;\\
\supit{j}Institut dÕAstrophysique de Paris, UMR7095 CNRS, Universit\'e Pierre \& Marie Curie, 98bis boulevard Arago, 75014 Paris, France;\\
\supit{k}Centro de Astrof\'isica, Universidade do Porto, Rua das Estrelas, 4150-762 Porto, Portugal;\\
\supit{l}Laborat\'orio Nacional de Astrof\'isica (LNA/MCTI), Rua Estados Unidos, 154 Itajub\'a - MG, Brazil;\\
\supit{m}Departamento de F\'isicaÐICExÐUFMG, Av. Ant\^onio Carlos, 6627, 30270-901, Belo Horizonte, MG, Brazil;\\
\supit{n}Departamento de F\'isica Te\'orica e Experimental, Universidade Federal do Rio Grande do Norte, Natal (RN), Brazil 
\end{small}
}

\authorinfo{Further author information: Send correspondence to E.A., artigau@astro.umontreal.ca, Telephone: 1 514-343-6111, ext. 3190}

\begin{document} 
  \maketitle 

\begin{abstract}
SPIRou is a near-IR \'echelle spectropolarimeter and high-precision velocimeter under construction as a next-generation instrument for the Canada-France-Hawaii-Telescope. It is designed to cover a very wide simultaneous near-IR spectral range (0.98-2.35 $\mu$m) at a resolving power of 73.5\,K, providing unpolarized and polarized spectra of low-mass stars at a radial velocity (RV) precision of 1\,m/s. The main science goals of SPIRou are the detection of habitable super-Earths around low-mass stars and the study of how critically magnetic fields impact star / planet formation.  Following a successful final design review in Spring 2014, SPIRou is now under construction and is scheduled to see first light in late 2017. We present an overview of key aspects of SPIRou's optical and mechanical design.
\end{abstract}



\section{INTRODUCTION}
\label{sec:intro}  
The {\it Spectro-Polarimetre Infra-Rouge} (SPIRou), is a state-of-the-art near-infrared spectro-polarimeter optimized for the detection of exoplanets using the radial velocity (RV) technique, and for the detection of magnetic fields in young embedded stellar objects.  SPIRou is currently under construction for first light at the CFHT in 2017 following successful final design review in spring 2014. SPIRou builds upon the success of three instruments, the optical spectropolarimeter ESPaDOnS\cite{Donati:2006}, which pioneered the exploration of large-scale magnetic topologies in T Tauri stars and is currently in use at CFHT, the radial-velocity (RV) spectrograph SOPHIE\cite{Bouchy:2006} (OHP) and the RV spectrograph HARPS\cite{Pepe:2000a} (La Silla, 3.6-m telescope) that revolutionized radial-velocity (RV) planet-searches by being the first instrument to reach 1\,m/s RV accuracy. A summary of the main instrumental caracteristics of SPIRou is given in table~\ref{tbl1}. More than a decade after HARPS commissioning, only a handful of other facilities have reached this impressive stability level in the optical ($400-700$\,nm), and none, yet, in the near-infrared domain ($1-2.4\mu$m).

\begin{table}[!htbp]
\caption{Key instrumental characteristics of SPIRou} 
\label{tbl1}
\begin{center}       
\begin{tabular}{|p{0.35\linewidth}|p{0.5\linewidth}|} 
\hline
Overall design type & Fiber-fed, bench-mounted, double-pass, cross-dispersed spectropolarimeter \\  \hline
Spectral Range & $0.978 - 2.437\mu$m (optimized for $0.978 - 2.363\mu$m)\\  \hline
Radial-velocity stability & $<1$\,m/s\\ \hline
Resolution & $\lambda/\Delta\lambda > $70\,000\\ \hline
Faint-target sensitivity & $H=14$, $10\sigma$, 30\,min\\ \hline
Achromatic polarimeter & $<1$\% cross-talk\\\hline
Throughput (Optics) & All wavelengths $> 45$\%, Average $> 50$\%\\ \hline
Working temperature & 80\,K \\ \hline
Science array & H4RG-15\footnote{http://www.teledyne-si.com/imaging/hawaii4rg.html} HgCdTe array, $4096 \times 4096$, 15\,$\mu$m pixels\\ \hline
\end{tabular}
\end{center}
\end{table}		
	
\begin{figure}
   \begin{center}
   \begin{tabular}{c}
   \includegraphics[height=7cm]{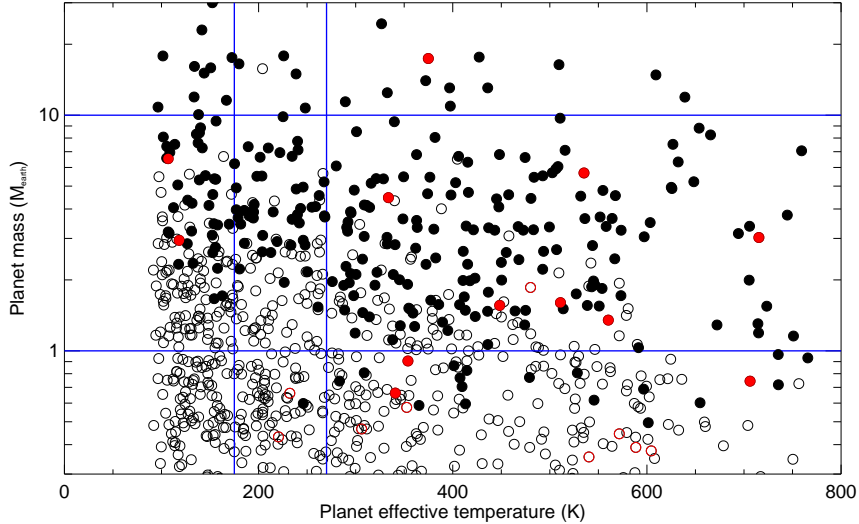}
   \end{tabular}
   \end{center}
   \caption[example] 
   { \label{fig_survey} Expected planet yield from a SPIRou RV planet search over 300 nights and including 360 mid-to-late M dwarfs. Filled circled indicate detected planets, open circles undetected ones and red circles (both filled and open) represent transiting planets. A rough indication of the limits of the habitable zone, both in mass and temperature, is indicated by blue lines. Most planets with $>2$M$_{\oplus}$ in the habitable zone are detected. Interestingly, a sample of sub-M$_{\oplus}$  planets with $>350$\,K is also predicted to be uncovered. Whether this population of planets exists remains to be seen, but this could be an interesting side-product of the RV search.  }
   \end{figure}

SPIRou will be a pioneer for RV planet searches in the infrared, rather than in the visible, a paradigm shift that enables targeting low-mass stars, around which Earth-sized planets are much easier to detect than around the more massive stars accessible to spectrographs operating in the visible. SPIRou will reach an RV accuracy of better than 1 m/s, sufficient to detect the reflex motion induced by Earth-size planets orbiting around mid-M dwarfs. SPIRou will be used to monitor the RV of several hundreds of low-mass stars over a period of 3 to 5 years to discover hundreds of new terrestrial exoplanets, a sizeable fraction of which will be in the so-called habitable zone (HZ), this region around a star warm enough to sustain liquid water on the surface of a planet. SPIRou will determine, for the first time, the fraction of low-mass stars harboring habitable planets. Figure~\ref{fig_survey} shows mass/temperature diagram for expected planet discoveries from the SPIRou RV survey. Second, concomitant with SPIRou, NASA will launch the Transiting Exoplanet Survey Satellite (TESS\cite{Ricker:2010}) in 2017 to conduct an all-sky survey over two years to find new and nearby transiting exoplanets. TESS is expected to find thousands of new transiting exoplanet candidates, including several hundreds of super-Earths, the majority around low-mass stars. Transiting planets are rare, but very important, cases of planets that happen to pass, from our perspective, in front of their star once every orbit, enabling a direct measurement of the planet radius. Confirmation of TESS' planets will require a mass measurement that can only be obtained through IR precision velocimetry with an instrument like SPIRou, a capability that does not exist yet. SPIRou will be an essential ancillary facility to the TESS mission. 
	
Towards the end of this decade, TESS and SPIRou together will have assembled a sample of well characterized Earths and Super-Earths orbiting in the HZ of low-mass stars. This will set the stage for the James Webb Space Telescope (JWST), slated for launch in 2018, to probe the atmosphere of those habitable words through transit spectroscopy, to search for water and even biomarkers. The NIRISS\cite{Doyon:2012} instrument aboard JWST features a mode specifically designed for high-precision transit spectroscopy. JWST will be humanity's first opportunity to detect life outside the solar system.

The second main goal is to explore the impact of magnetic fields on star and planet formation, by detectingby detecting and characterizing the large-scale magnetic fields of young stellar objects, especially embedded (class I) protostars that are mostly out of reach of existing optical spectropolarimeters like ESPaDOnS, as well as T Tauri stars (class II and III protostars) and FU-Ori like protostellar accretion discs. SPIRou will also be able to tackle many other exciting research topics in stellar physics (e.g., dynamo of fully convective stars, weather patterns at the surfaces of brown dwarfs\cite{Crossfield:2014}), in planetary physics (e.g., winds and chemistry of solar-system planet atmospheres\cite{Widemann:2008a}) galactic physics (e.g., stellar archaeology) as well as in extragalactic astronomy.

\begin{figure}
   \begin{center}
   \begin{tabular}{c}
   \includegraphics[width=12cm]{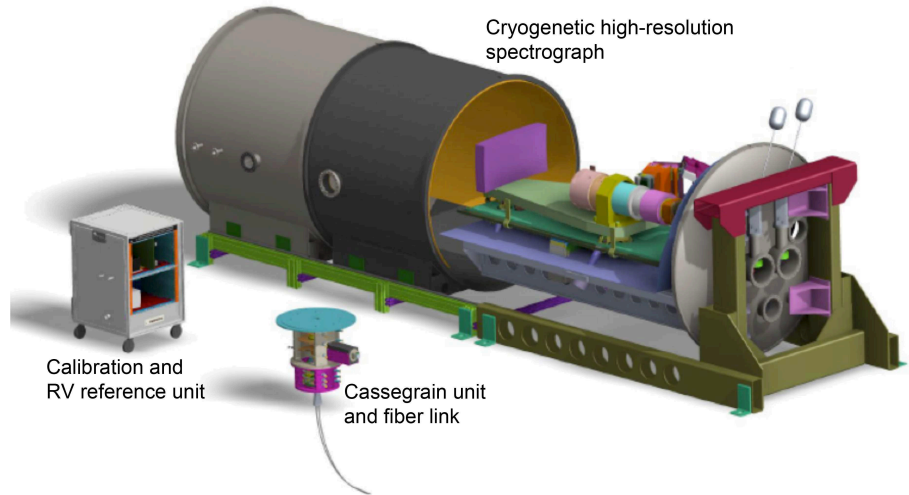}\\
   \end{tabular}
   \end{center}
   \caption[example] 
   { \label{fig_global}  Schematic view of the SPIRou instrument main components. The Cassegrain unit contains polarimetric analyzers injecting star light into two 35\,m-long fibers. The main spectrograph, to be installed at the CFHT coud\'e room, consists of a 3m-long cryostat (see also figure\,\ref{fig_cryostat}). A calibration unit can inject light form a stabilized etalon and ThAr lamps into the science path. }
   \end{figure}

 This contribution provides a brief overview of the science goals of SPIRou, and summarizes key aspects of its optical and mechanical design. Figure~\ref{fig_global} shows a schematic view of SPIRou's three main components that are described in this contribution: the Cassegrain unit for polarimetric analysis and fiber injection (section~\ref{cassegrain}), the bench-mounted  spectrograph (section~\ref{spectrograph}) and the calibration unit that provide long-term monitoring of SPIRou's RV stability (section~\ref{calibration}). Much more detailed accounts of the foreseen contribution of SPIRou to various fiels of astrophysics can be found in Artigau, Donati \& Delfosse (2011)\cite{Artigau:2011}, Delfosse et al. (2013)\cite{Delfosse:2013}. Santerne et al (2013)\cite{Santerne:2013a} explores in depth the synergies between SPIRou and space-based missions. Aspects not covered here include but that have received significant attention trough the project history include the command and control\cite{Barrick:2012} as well the data reduction and simulation tools\cite{Artigau:2012}. The complete science case of SPIRou is available online\footnote{http://spirou.irap.omp.eu/Science2}.

\section{Cassegrain unit}
	 \label{cassegrain}
\begin{figure}
   \begin{center}
   \begin{tabular}{c}
   \includegraphics[width=15cm]{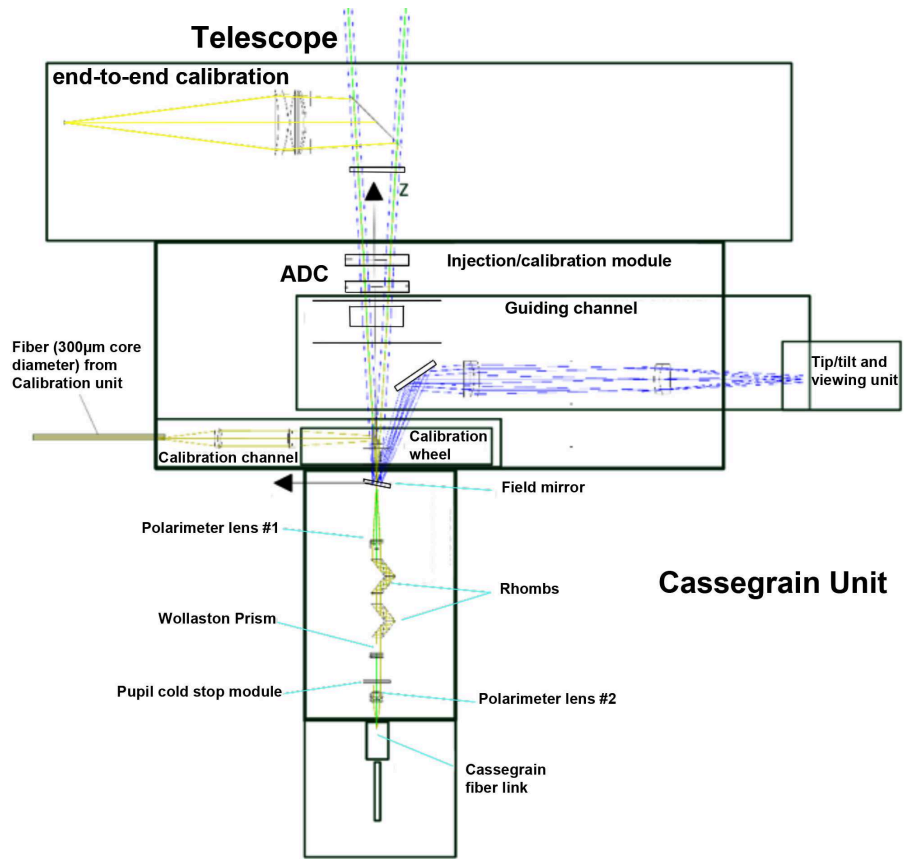}
   \end{tabular}
   \end{center}
   \caption[example] 
   { \label{fig_cassegrain} Optical layout of the SPIRou cassegrain Unit}
   \end{figure}

The Cassegrain unit is composed of two main modules; figure~\ref{fig_cassegrain} shows a schematic view of this unit. This section provides an updated overview of the Cassegrain unit; a more in-depth accounting is provided in Pares et al. (2012)\cite{Pares:2012}. The first module is dedicated to the injection and input from calibration light into the science channel. This module includes an an atmospheric dispersion corrector (ADC) which provides correction for an airmass range of $z=1-2.5$ over the entire near-infrared domain. The second element of that module stabilizes the field of view to minimize near-field drifts at the level of the flux injection into the science channel fiber. This modules has an SWIR camera for field viewing and guides on the reflection from the field mirror.  The field mirror is a parallel window inclined at 10$^\circ$, it features an antireflection coating on the upper side (toward telescope) and a mirror coating with a circular aperture on the lower (toward the polarimeter module). The mirror coating reflects the observed field of view toward the guiding channel. This permits the identification of the field of view and proper acquisition of the science target. The target is centered on the circular aperture of the field mirror. During long exposures, the reflection of the wings of the target's point-spread function and the Fresnel reflection of the circular aperture ensure that the target is correctly centered in the circular hole of the field mirror. Short-time exposures done during tip-tilt correction use the flux from Fresnel reflection to analyze the tip-tilt displacements of the star. The corrections are sent to the image stabilization unit (ISU). The ISU is a thick fused silica plate. The field is stabilized by tilting the ISU plate through a closed-loop.

The last element of the module is the calibration channel with the calibration wheel (prisms, linear polarizers) and optics for converting the F/4 beam from the RV reference fiber to an F/8 beam. The calibration wheel redirects the light from the calibration and RV module on the optical axis of the telescope toward the polarimeter module. Polarimetric calibration is done with linear polarizers installed in the calibration wheel. During science observations, the calibration wheel is positioned with a clear aperture large enough as not to block the light toward the guiding channel.

The second module in the Cassegrain unit is the polarimeter. It contains a field mirror that reflects part of the light to the guiding camera and defines the field stop. In the polarimeter module, lenses convert the F/8 beam from the telescope to an F/4 beam at fiber bundle level. These lenses optically conjugate the field mirror circular aperture and the entrance of the fibers. The beam between the lenses is collimated. The polarimeter itself is composed of a first lens (polarimeter lens \#1), two  quarter-wave rhombs, a Wollaston polarization analyzer, a pupil cold stop and finally the exit optics (polarimeter lens \#2) injecting the light into the Cassegrain fiber link\cite{Micheau:2012}. At this stage, the two polarizations have been split into two independent fibers. The polarization analysis is done in the polarimeter module by two $\lambda$/4 retardation plates and a Wollaston working in parallel beam. The retardation plates are modified Fresnel $\lambda$/4  rhombs that do not introduce beam deviation. By changing the orientation of the rhombs, the Stokes parameters science target are determined. At the end of the polarimeter module, the light is separated in two beams with orthogonal polarizations by the Wollaston. A pupil cold stop module, placed after the Wollaston prism, blocks the light from the baffles of the telescope to reduce the thermal background. The thermal background of the Cassegrain unit is reduced by use of low emissivity screens, placed between each optical element.  These two beams are then injected in two fibers separated by 250\,$\mu$m. These fluoride fibers carry the polarization-analyzed light between the Cassegrain unit and the main cryostat spectrograph and its pupil slicer.

\clearpage
\section{Fluoride fibers and pupil slicer}
This part of the optical train is comprised of two sub-systems. The first one is the fiber link consisting of the set of optical fluoride fibers linking the Cassegrain unit, the calibration module and the spectrograph. The second sub-system is the pupil slicer that optically dissects the science fibers (in pupil plane) into 4 slices that form the entrance slit of the spectrograph. The slicer and associated tools needed for alignment constitute a sub-system of the cryogenic spectrograph. The long (fluoride) fibers are made of a special optical material made of ZrF4 that enables good transmission beyond $2.0 \mu$m ($K$ band). Transmission into the $K$-band is a unique observing capability for SPIRou; no other infrared precision radial velocity spectrometer will operate beyond $1.8 \mu$m as they all are limited to use off-the-shelf Si fibers that have poor transmission into the $K$ band. The long (35\,m) optical fibers requires very low-attenuation  ($<13$\,db/km) to meet the SPIRou throughput requirement. Such low attenuation is achieved through a sophisticated (and somewhat expensive) purification process of the fiber material. Inside the spectrograph, after custom-made hermetic connectors, the circular fluoride fibers are coupled to a 1.4\,m-long segment of octagonal fibers that feeds the pupil slider. The purpose of the octagonal fiber is to scramble light and minimize systematic RV residual effects associated with potential non-uniform illumination within the science fibers. Experience has shown that octagonal fibers are very effective at improving the accuracy of precision radial velocity spectrographs\cite{Perruchot:2011}.

\begin{figure}
   \begin{center}
   \begin{tabular}{c}
   \includegraphics[width=12cm]{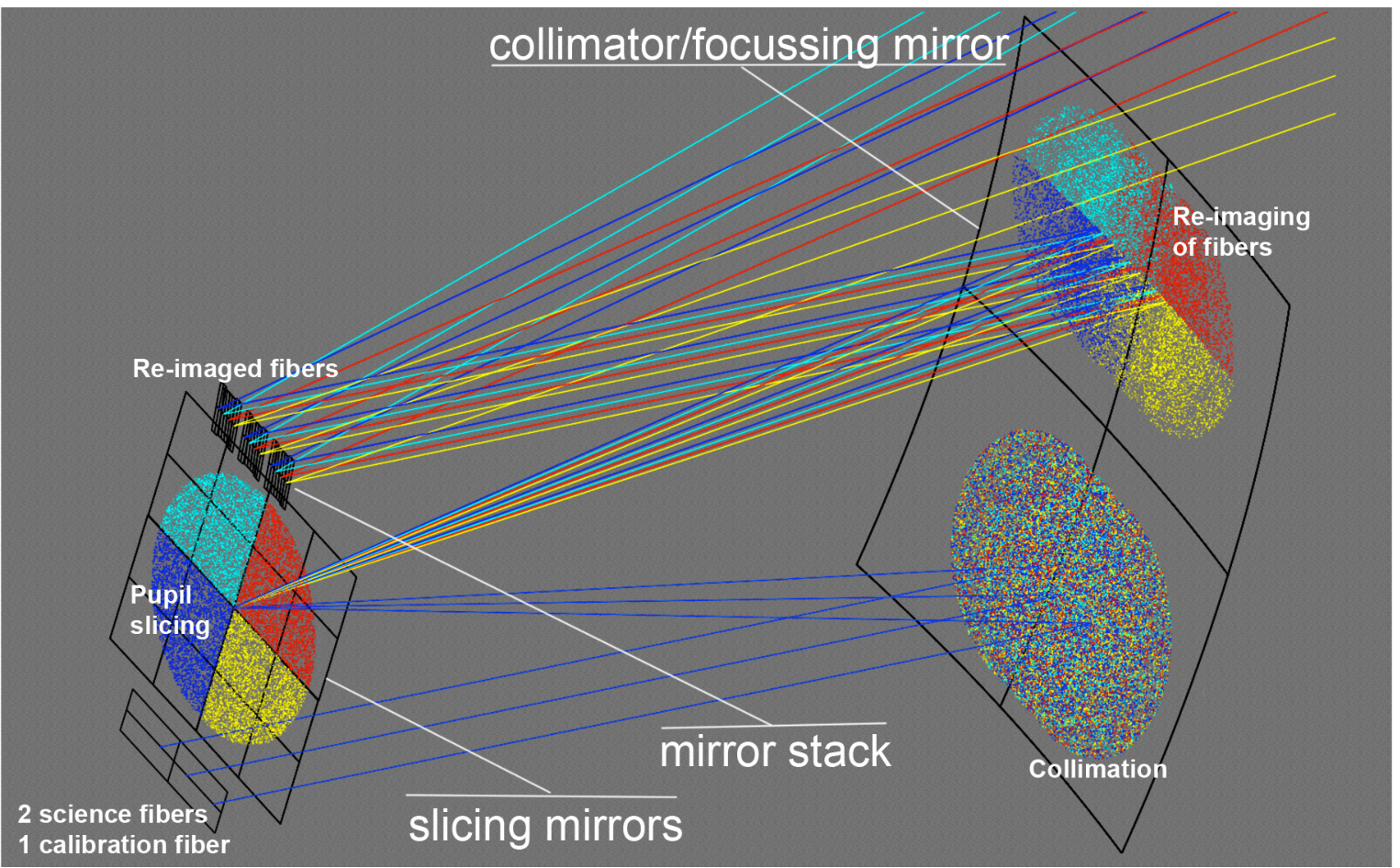}\\
   \includegraphics[width=12cm]{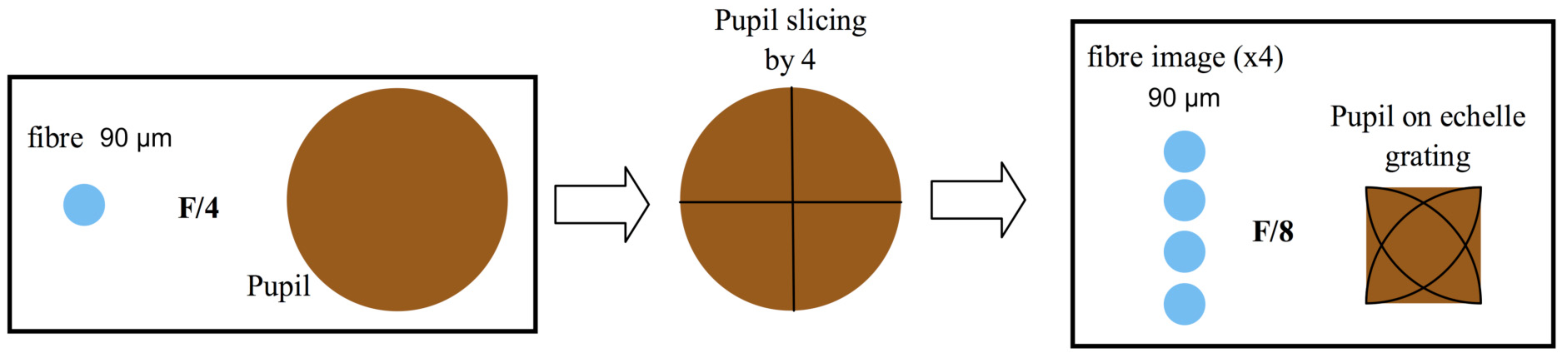}
   \end{tabular}
   \end{center}
   \caption[example] 
   { \label{fig_slicer} (top) Schematic view of the pupil slicer. The output from the science and RV reference fiber is collimated onto a pupil-slicing mirror consisting of four flat surfaces tilted relative to each other. The pupil images are reflected back onto the collimating mirror to be re-imaged into fiber images. (bottom) Cartoon view of the pupil slicing principle. }
   \end{figure}

The output beam of these two fibers is F/4 with a 90\,$\mu$m diameter which is transformed into an F/8 beam, before transfer to the spectrograph. The use of a simple focal extender would have increased the image size and degraded the resolution, significantly hampering the scientific output of SPIRou and its radial velocity accuracy. The solution to obtain the required slit and pupil profile on the R2 grating is to use a pupil slicer. The adopted pupil slicer design is summarized in figure~\ref{fig_slicer}. SPIRou's pupil slicer is based on free space optical components. To have the maximum stability in the cryogenic chamber, the optical design is only composed of mirrors. Collimation of light coming from fibers is done by the {\it collimator/focusing mirror} with a parabolic surface. It creates a circular, 3.125\,mm diameter, pupil onto the {\it slicing mirrors}. Slicing mirrors are 4 identical flat rectangular mirrors with different $y$-tilt to form 12 images along a slit (4 images for each fiber) at the {\it mirror stack} level after reflection on the {\it collimator/focusing mirror}. 

These images are then reflected with the correct angle by the mirror stack to obtain the pupil profile required on the R2 grating. The mirror stack is composed of 9 flat mirrors mounted together and with different tilts. Considering a space of 20\,$\mu$m between two slices within the slit, each flat mirror will have a width of 110\,$\mu$m.
	
\clearpage

\section{Spectrograph}
 \label{spectrograph}
The main science goals of SPIRou, the discovery of terrestrial planets in the habitable zone of M dwarfs and spectropolarimetry of young stellar objects, lead to strong requirements on the design of SPIRou. The radial velocity stability of SPIRou over its useful live must be smaller than the typical RV shift induced by an habitable exoplanet around an M dwarf of $\sim$1\,m/s, setting the bar to SPIRou's RV accuracy and many of the adopted engineering solutions.

\begin{figure}
   \begin{center}
   \begin{tabular}{c}
   \includegraphics[width=16cm]{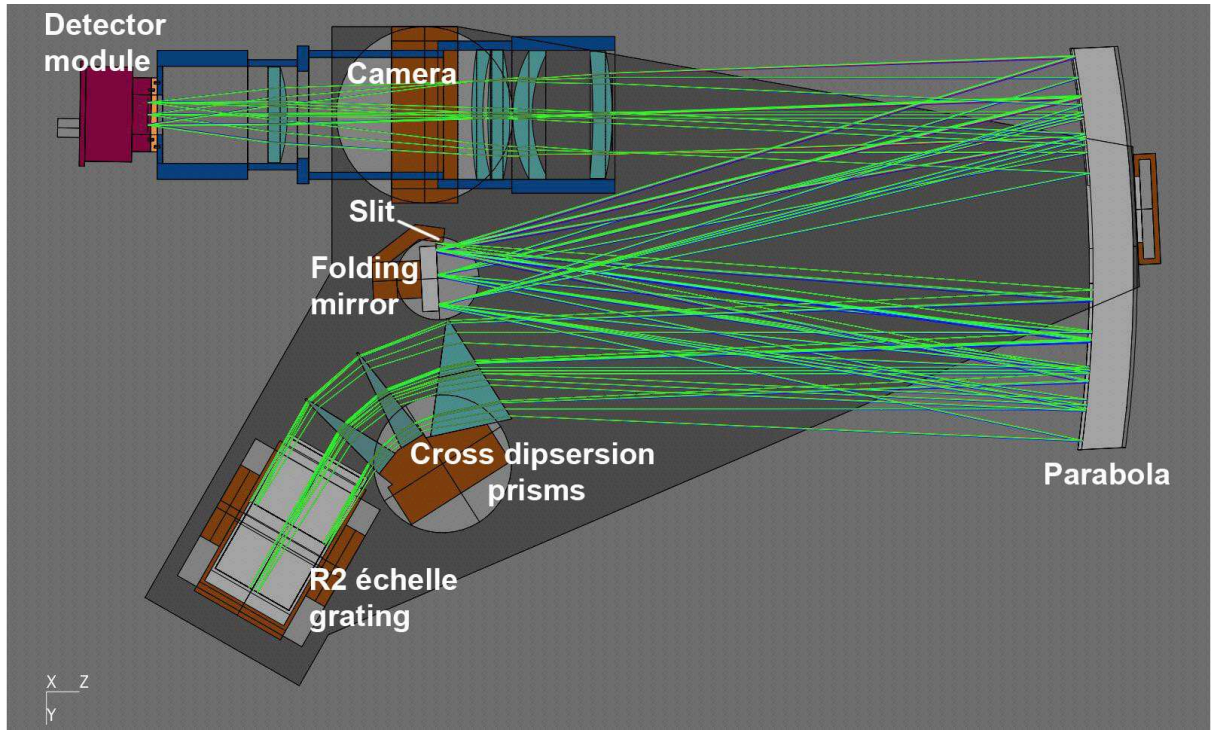}
   \end{tabular}
   \end{center}
   \caption[example] 
   { \label{fig_raytrace} schematic view of the spectrograph optical train. The use of double-pass cross-dispersing prisms and a fold mirror lead to a relatively compact design. }
   \end{figure}

The adopted SPIRou spectrograph design is a near-infrared fiber-fed double-pass cross-dispersed \'echelle spectrograph, based on ESPaDOnS optical design (see figure~\ref{fig_raytrace}). It includes one parabola, an \'echelle grating, a train of cross-dispersering prisms (in double pass), a flat folding mirror, a refractive camera and a detector. We provide here a summary of the final design of SPIRou spectrograph, earlier versions having, most notably, explored both R2 and R4 grating options\cite{Thibault:2012}. The former has been selected for construction after considering the significant risk involved in the development of the R4 grating.

The optical path sequence is the following:
\begin{itemize}
\item The parabola collimates the fiber fed pupil slicer image slit beam;
\item The collimated beam goes through the cross disperser prisms (first pass);
\item The echelle grating diffracts the collimated beam;
\item The diffracted collimated beam goes back through the cross disperser prisms (second pass);
\item The parabola focuses the diffracted collimated beam (second pass);
\item The flat mirror folds the diffracted focused beam;
\item The parabola collimates the diffracted focused beam;
\item The refractive camera focuses the diffracted collimated beam on the detector.
\end{itemize}

\subsection{Camera and lens mounting}
The optical bench is made of 5 lenses. The lenses mounting technique, described below, is the same the one used for the camera WIRCAM\cite{Puget:2004}. Each lens is held in a precisely machined aluminum cell. The lens is held in the cell axially with  beryllium-copper leaf springs SPIRou so that the total force applied to the lens is approximately 3 times its weight. An aluminum ring is located between the springs and the lens so that the springs don't come in contact with the lens. The lens is held radially with three T-shaped nylon or Teflon pads evenly distributed on the perimeter. The length of pads is adjusted so that the gap between the pad and the lens is close to zero at room temperature and at operating cryogenic temperature. Spring washers are used to compensate the differential thermal contraction between screws (stainless steel) and pads (nylon). 

   		   \begin{figure}
   \begin{center}
   \begin{tabular}{c}
   \includegraphics[width=16cm]{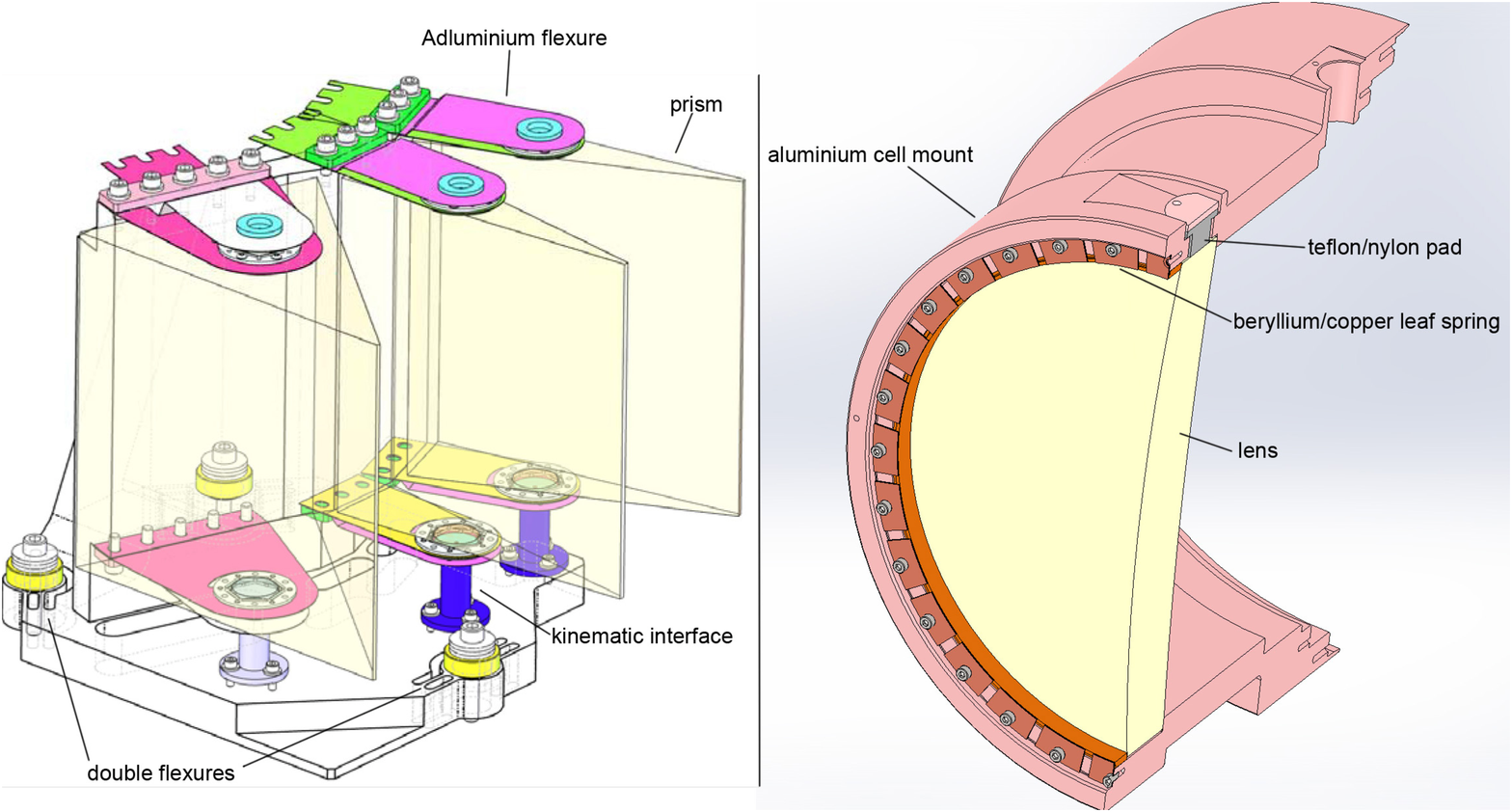}
   \end{tabular}
   \end{center}
   \caption[example] 
   { \label{fig_mounts} Left : View of the prism mounting scheme; a similar scheme is used for the grism. The prisms are constrained in all but vertical dimension through aluminum flexures glued to their surface. The flexure include a stress relieve pattern at the level of the glue point. Vertically, prisms are constrained through a flat/ball/fall kinematic interface\cite{Reshetov:2012}. Right: Cutoff view of the lens mounting scheme; lenses are constrained radially through nylon or Teflon pads which length has been adjusted to compensate differential thermal contraction between the lens's glass and the aluminum cell. Axially, lenses are held in place through beryllium/copper leaf springs.}
   \end{figure} 

Studies on the project WIRCAM have shown that a maximum gradient of 20\,K can be expected between the lens and the barrel during cooldown. In the case of WIRCAM, a cool down time of 24\,h was used to make the calculation. For WIRCAM, models and calculations showed a 10-fold safety margin when using the pad scheme discussed above. In the case of SPIROU, a cool down time of 3 days is expected. The cool down time being even slower than for WIRCAM, we are very confident that the method presented above will be safe for SPIRou.  

\subsection{Detector choice}
The broad simultaneous wavelength coverage of SPIRou combined with its high resolution implies that at least $\sim$$100 000$ resolution elements need to be recorded. Considering the adopted pupil slicer design that produces 4 slit images per science channel, and that the fiber image is $\sim$2 pixels in diameter, at least 36 pixels will be illuminated for each resolution element (4 images per science fiber, 1 for the calibration channel, leading to 9 images, each consisting of $2\times2$ pixels). This leads to the illumination of $\sim$4$\times 10^6$ pixels. Considering an inter-image and inter-order gap of $\sim$2\,pixels and the trapezoid shape of order placement, a science array with at least $10^7$ pixels is required. The H4RG\cite{Blank:2012} detector developed by Teledyne is the best available detector for this purpose, with $4096 \times 4096$ pixels and $<5$e$^-$ effective readout noise after multiple readouts. Orders only occupy $\sim$$2500$ of the 4096 pixels height of the science array, leaving significant leeway should the selected science array present cosmetic defects on one of its sides. Provision to move the science array has been done in the detector attachement to the camera.

\begin{figure}[!h]
\begin{center}
\begin{tabular}{c}
\includegraphics[width=8cm]{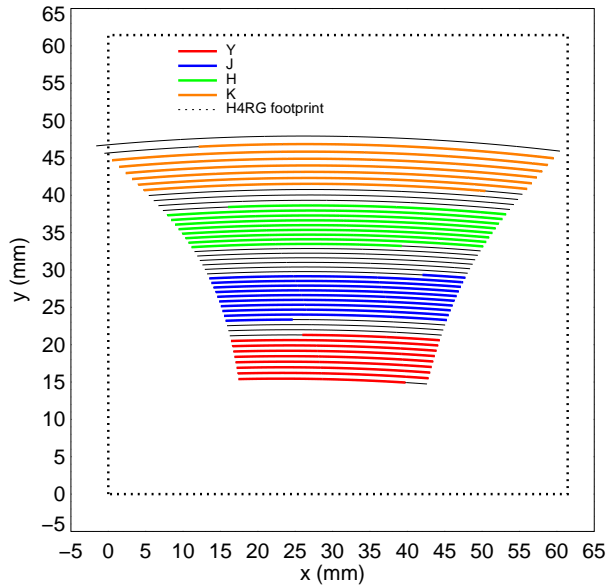}
\end{tabular}
\end{center}
\caption[example] 
   { \label{fig_h4rg} Order disposition on SPIRou's science array. The spectral orders at the red end of domain cover the entire width of the science array. Vertically, the orders only cover $\sim$$2500$ pixels, leaving leeway to move the array up or down depending on its cosmetics.}

\end{figure} 

\subsection{Cryostat}

\begin{figure}
\begin{center}
\begin{tabular}{c}
\includegraphics[width=12cm]{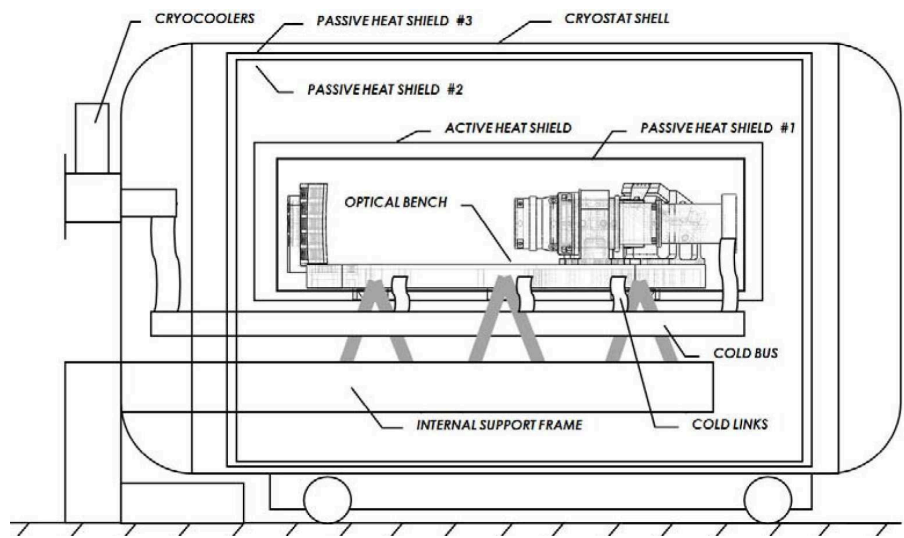}
\end{tabular}
\end{center}
\caption[example] 
   { \label{fig_cryostat} 
    Schematic view of the main elements of SPIRou's cooling and thermal control system. The optical bench is encased wihtin a first passive shield, which is itself encased in an active shield. The optical bench and these two shields lay on a cold bus linked to cryocoolers, and are supported by the internal support frame through G10 rods. Two more passive head shields enshroud the entire instrument, providing a global thermal stability at the 1\,mK level for the optical. }
\end{figure}

The SPIRou cryostat has a cylindrical form with an outer diameter of 1.7\,m by a 3.3\,m length in operational mode (when closed). The cylindrical portion of the cryostat slides along its axis to allow easy access to the optical bench (see figures~\ref{fig_global} and \ref{fig_cryostat}). When opened, the cryostat occupies a footprint of $5.7 \times 1.7$\,m, and is 1.8\,m high. The total mass of the cryostat estimated to be 3600\,kg. This number excludes vacuum pumps and cryogenic pumps, electronics and other supporting hardware. The optical bench is located in a horizontal position. It is supported at three points by a hexapod type arrangement from an internal warm support frame (see figure~\ref{fig_cryostat}). The hexapods have high thermal resistance to insulate the optical bench from the warm internal frame. The internal support frame is cantilevered from the stationary end of the cryostat. For the ease of maintenance all instrument feed-through and cryo-coolers are located on the stationary parts of the cryostat. In total there are four heat shields: two passive heat shields attached to the cryostat shell, one active shield which is suspended from the internal frame, and one last passive shield, which is attached to the inside of the active shield.

\clearpage

\section{Calibration unit }
\label{calibration}
The calibration unit goal is to calibrate and to characterize the spectrometer response to secure the highest possible RV stability, both for short term (one night) and long term (several years) activity. Therefore the calibration unit has to provide calibration sources in order to perform the following calibrations: 1) localization and geometry of spectral orders; 2) determination of the flat field (pixel response) and blaze profile response; 3) determination of the wavelength calibration; 4) determination of the radial velocity reference (zero point). Furthermore, a cold source with low thermal emissivity is required in case of no simultaneous calibration as not to add an additional thermal background to during science observations.

The calibration light feeds the spectrograph through two channels: one following the stellar beams path through the Cassegrain Unit, the other directly at the spectrograph slicer entrance.

The unit architecture for SPIRou comes from HARPS and SOPHIE heritage. Some adaptations are made to be compliant with the SPIRou requirements.

The calibration module is composed of 3 parts:
\begin{itemize}
\item Calibration unit
\item Radial Velocity Reference unit (RV reference unit) 
\item RV reference fiber
\end{itemize}

The calibration unit provides light from different selectable sources to feed two RV reference fiber, one linked to the Cassegrain unit and the other to the spectrograph through the slicer unit.

The RV reference unit is essentially a Fabry-Perot etalon housed in a temperature-controlled vacuum enclosure. The etalon is fiber-fed by a bright white lamp. A symmetrical set-up of parabolas couples the input fiber to the exit fiber, the etalon being located in the collimated beam between the two parabolas. The output fiber is connected to the calibration module. The RV reference unit is considered as a light source of the calibration module. This unit is provided by Observatory of Geneva.

The RV reference fiber package includes the two fiber links:

\begin{itemize}
\item The Cassegrain calibration channel: links directly the calibration module to the Cassegrain unit.
The calibration beams follow the same optical path as the science beams;
\item The Reference calibration channel: links directly the calibration module to the spectrometer through the slicer unit.
\end{itemize}

\bibliography{bibdesk}

\begin{thebibliography}{1}
 
\bibitem{Donati:2006}{Donati}, J.-F., {Catala}, C. \& {Landstreet}, J.~D., and {Petit}, P., ``{ESPaDOnS: The New Generation Stellar Spectro-Polarimeter. Performances and First Results},'' in [{\em Astronomical Society of the Pacific Conference Series}{\nolinebreak\hspace{0.1em}]},  {Casini}, R. and {Lites}, B.~W., eds., {\em Astronomical Society of the Pacific Conference Series}, {\bf 358},  362 (2006).
\bibitem{Bouchy:2006}{Bouchy}, F. and {Sophie Team}, ``{SOPHIE: the successor of the spectrograph ELODIE for extrasolar planet search and characterization},'' in [{\em Tenth Anniversary of 51 Peg-b: Status of and prospects for hot Jupiter studies}{\nolinebreak\hspace{0.1em}]},  {Arnold}, L., {Bouchy}, F., and {Moutou}, C., eds.,  319--325 (2006).
\bibitem{Pepe:2000a}{Pepe}, F., {Mayor}, M., {Delabre}, B. et al, ``{HARPS: a new high-resolution spectrograph for the search of extrasolar planets},'' {\em Proc. SPIE} {\bf 4008},  582--592 (2000).
\bibitem{Ricker:2010}{Ricker}, G.~R., {Latham}, D.~W., {Vanderspek}, R.~K. et al, ``{Transiting Exoplanet Survey Satellite (TESS)},'' in [{\em American Astronomical Society, Meeting Abstracts \#215}{\nolinebreak\hspace{0.1em}]},  {\em Bulletin of the American Astronomical Society}, {\bf 42},  450.06 (2010).
\bibitem{Doyon:2012}{Doyon}, R., {Hutchings}, J.~B., {Beaulieu}, M. et al, ``{The JWST Fine Guidance Sensor (FGS) and Near-Infrared Imager and Slitless Spectrograph (NIRISS)},'' {\em Proc. SPIE} {\bf 8442} (2012).
\bibitem{Crossfield:2014}{Crossfield}, I.~J.~M., {Biller}, B., {Schlieder}, J.~E. et al, ``{A global cloud map of the nearest known brown dwarf},'' {\em \nat}~{\bf 505},  654--656 (2014).
\bibitem{Widemann:2008a}{Widemann}, T. \& {Lellouch}, E., and {Donati}, J.-F., ``{Venus DopplerWinds at Cloud Tops Observed with ESPaDOnS at CFHT},'' in [{\em European Planetary Science Congress 2008}{\nolinebreak\hspace{0.1em}]},   561 (2008).
\bibitem{Artigau:2011}{Artigau}, {\'E}. \& {Donati}, J.-F., and {Delfosse}, X., ``{Planet Detection, Magnetic Field of Protostars and Brown Dwarfs Meteorology with SPIRou},'' in [{\em 16th Cambridge Workshop on Cool Stars, Stellar Systems, and the Sun}{\nolinebreak\hspace{0.1em}]},  {Johns-Krull}, C., {Browning}, M.~K., and {West}, A.~A., eds., {\em Astronomical Society of the Pacific Conference Series}, {\bf 448},  771 (2011).
\bibitem{Delfosse:2013}{Delfosse}, X., {Donati}, J.-F., {Kouach}, D. et al, ``{World-leading science with SPIRou - The nIR spectropolarimeter / high-precision velocimeter for CFHT},'' in [{\em SF2A-2013: Proceedings of the Annual meeting of the French Society of Astronomy and Astrophysics}{\nolinebreak\hspace{0.1em}]},  {Cambresy}, L., {Martins}, F., {Nuss}, E., and {Palacios}, A., eds.,  497--508 (2013).
\bibitem{Santerne:2013a}{Santerne}, A., {Donati}, J.-F., {Doyon}, R. et al, ``{Characterizing small planets transiting small stars with SPIRou},'' in [{\em SF2A-2013: Proceedings of the Annual meeting of the French Society of Astronomy and Astrophysics}{\nolinebreak\hspace{0.1em}]}, {Cambresy}, L., {Martins}, F., {Nuss}, E., and {Palacios}, A., eds., 509--514 (2013).
\bibitem{Barrick:2012}{Barrick}, G.~A., {Vermeulen}, T., {Baratchart}, S. et al, ``{SPIRou @ CFHT: design of the instrument control system},'' {\em Proc. SPIE} {\bf 8451} (2012).
\bibitem{Artigau:2012}{Artigau}, {\'E}., {Bouchy}, F. et al, ``{SPIRou @ CFHT: data reduction software and simulation tools},'' {\em Proc. SPIE} {\bf 8451} (2012).
\bibitem{Pares:2012}{Par{\`e}s}, L., {Donati}, J.-F., {Dupieux}, M. et al, ``{Front end of the SPIRou spectropolarimeter for Canada-France Hawaii Telescope},'' {\em Proc. SPIE} {\bf 8446} (2012).
\bibitem{Micheau:2012}{Micheau}, Y., {Bouchy}, F., {Pepe}, F. et al, ``{SPIRou @ CFHT: fiber links and pupil slicer},'' {\em Proc. SPIE} {\bf 8446} (2012).
\bibitem{Perruchot:2011}{Perruchot}, S., {Bouchy}, F., {Chazelas}, B. et al, ``{Higher-precision radial velocity measurements with the SOPHIE spectrograph using octagonal-section fibers},'' {\em Proc. SPIE} {\bf 8151} (2011).
\bibitem{Thibault:2012}{Thibault}, S., {Rabou}, P., {Donati}, J.-F. et al, ``{SPIRou @ CFHT: spectrograph optical design},'' {\em Proc. SPIE} {\bf 8446} (2012).
\bibitem{Puget:2004}{Puget}, P., {Stadler}, E., {Doyon}, R. et al, ``{WIRCam: the infrared wide-field camera for the Canada-France-Hawaii Telescope},'' {\em Proc. SPIE} {\bf 5492},  978--987 (2004).
\bibitem{Reshetov:2012}{Reshetov}, V., {Herriot}, G., {Thibault}, S., {D{\'e}saulniers}, P. \& {Saddlemyer}, L., and {Loop}, D., ``{Cryogenic mechanical design: SPIROU spectrograph},'' {\em Proc. SPIE} {\bf 8446} (2012).
\bibitem{Blank:2012}{Blank}, R., {Beletic}, J.~W., {Cooper}, D. et al, ``{Development and production of the H4RG-15 focal plane array},'' {\em Proc. SPIE} {\bf 8453} (2012). \end{thebibliography}
\bibliographystyle{spiebib}   

\end{document}